# Experimental observation of the significant difference between surface and bulk Kondo processes in Kondo lattice YbCu$_2$Si$_2$

Yin-Zou Zhao,[1] Jiao-Jiao Song,[1] Qi-Yi Wu,[1] Hao Liu,[1] Chen Zhang,[1] Bo Chen,[1] Hong-Yi Zhang,[1] Zhen-Hua Chen,[2] Yao-Bo Huang,[2] Xue-Qing Ye,[1] Ya-Hua Yuan,[1] Yu-Xia Duan,[1] Jun He,[1] and Jian-Qiao Meng[1, *]

[1]*School of Physics, Central South University, Changsha 410083, Hunan, Peoples Republic of China*
[2]*Shanghai Institute of Applied Physics, CAS, Shanghai 201204, Peoples Republic of China*
(Dated: Sunday 8$^{\text{th}}$ October, 2023)

Synchrotron-based angle-resolved photoemission spectroscopy was employed to investigate the temperature evolution of the Yb 4$f$ spectral for surface and bulk in the Kondo lattice YbCu$_2$Si$_2$. Our study quantitatively distinguishes between the surface and bulk hybridization processes, revealing that the onset temperatures for both surface and bulk hybridization processes are significantly higher than the Kondo temperature. Additionally, we found that the effective surface Kondo temperature is much lower than that of the bulk. These findings offer valuable insights into the understanding of heavy fermion physics.

## I. INTRODUCTION

Heavy Fermion (HF), a typical strongly correlated electronic system, has rich and exciting physical phenomena, including antiferromagnetism, ferromagnetism, unconventional superconductivity, spin liquids, and topological states [1, 2]. These orders are intertwined, they compete, they cooperate, or they coexist. Singular physical properties of strongly correlated electronic systems tend to appear at the edge of the localized-itinerant transition. Uncovering the microscopic processes of the localized to the itinerant transition of the $f$ electrons in HFs is central to understanding their complex physical properties.

Earlier studies have suggested that coherence temperature $T^*$ marks the onset of collective hybridization between conduction electrons and localized $f$-electrons ($c$-$f$), leading to heavy quasiparticle bands at lower temperatures [3]. Generally, above $T^*$, the $f$ electrons can be described by a completely localized model, while well below $T^*$, the $f$ electrons can be described by an itinerant model [4]. However, recently, many angle-resolved photoelectron spectroscopy (ARPES) [5–11] and ultrafast optical spectroscopy [12–14] measurements have demonstrated that $c$-$f$ hybridization begins at high temperatures well above $T^*$. Notably, there is compelling evidence from ARPES to suggest that as the temperature continues to decrease, the $f$ electrons undergo relocalization, marking a transition at even lower temperatures [8–10].

Since the coordination numbers of the surface atoms differ significantly from those of the bulk atoms and ARPES is a surface-sensitive technique, surface states are observed in both Ce- [15, 16] and Yb-based [17–20] HFs. Different termination surfaces due to cleavage will produce different surface states [15–20]. In the case of Ce-based HF systems, the Ce 4$f$ spectral contributions from the surface and bulk regions occur at nearly the same binding energy [15, 16]. Conversely, Yb-based HF systems exhibit an intriguing feature: the surface and bulk-induced heavy quasiparticle states are energetically well separated [17–20]. This distinct behavior allows Yb-based HF materials to offer a unique opportunity for directly investigating both surface and bulk Kondo processes concurrently.

YbCu$_2$Si$_2$, a well-known intermediate valence compound and often regarded as a "hole" analog of the Ce-based HF systems, . It is a moderate HF system with a specific heat coefficient $\gamma \approx 150$ mJ/(K$^2$·mol)[21, 22]. Under ambient pressure, this material does not display magnetic order [23–25] or superconductivity [24]. De Haas-van Alphen (dHvA) measurements further indicate that as the temperature decreases, the localized Yb 4$f$ electrons transition towards itinerant behavior [21]. Under high pressure conditions, both resistivity [26, 27] and Mössbauer [27] measurements show evidence of ferromagnetic order at approximately 8 GPa. Remarkably, even under extreme pressures of up to 25 GPa, superconductivity remains conspicuously absent [26, 27], ostensibly quenched by the prevailing ferromagnetic state [28]. The Kondo temperature ($T_K$) of YbCu$_2$Si$_2$ is estimated to fall within the range of 35-60 K [21, 29, 30], encapsulating the complex interplay of electronic behaviors in this intriguing HF system.

In this study, we delve into the Yb 4$f$ electronic states of YbCu$_2$Si$_2$ utilizing high-resolution ARPES. The Yb 4$f$ photoelectron emission spectrum exhibits distinctive features characterized by a multi-peak structure spanning a wide range of binding energies. Specifically, two pronounced peaks emerge at approximately -1.3 eV and at the Fermi energy ($E_F$), originating from the $4f^{14} \to 4f^{13}$ transition. Additional multiple peaks manifest in the energy range spanning from -12 to -5.5 eV, attributed to the $4f^{13} \to 4f^{12}$ transition. We also observe two heavy quasiparticle bands situated at -0.53 and -1.83 eV, arising from the surface Yb layers. Most notably, our findings illuminate substantial disparities in the temperature dependence of Kondo processes between the surface and bulk. The bulk 4$f$ band remains discernible up to 280 K, demonstrating the persistence of Kondo processes at

2higher temperatures. Conversely, the surface $4f$ band exhibits a distinct behavior, diminishing around 145 K. This discrepancy in the temperature evolution of Kondo processes between surface and bulk contributes significantly to our understanding of the electronic behaviors in YbCu$_2$Si$_2$.

## II. EXPERIMENTAL DETAILS

High-quality single crystals of YbCu$_2$Si$_2$ were grown by an Sn self-flux method [21]. ARPES measurements were performed at the "Dreamline" beamline of the Shanghai Synchrotron Radiation Facility using a Scienta DA30 analyzer, and the vacuum was kept below $1\times10^{-10}$ mbar. All samples were cleaved *in situ* along the (001) plane at a low temperature of 10 K. We studied the $4f$ and conduction electron system. Deep electronic structures were probed with 220 eV. The Fermi surface (FS) topology of YbCu$_2$Si$_2$ along the $k_z$ (perpendicular) direction was mapped out using systematic photon energy dependence ($h\nu$ = 50-126 eV). Constant photon energies of 101 and 120 eV were used to probe the FSs ($k_x$-$k_y$ plane). Furthermore, time- and temperature-dependent measurements were carried out to gain a comprehensive understanding of the surface and bulk Yb $4f$ states. The typical angular resolution was ∼0.2°.

## III. RESULTS AND DISCUSSIONS

To elucidate the three-dimensional (3D) FS topologies of YbCu$_2$Si$_2$, a comprehensive investigation was conducted through photon energy-dependent normal emission measurements at a temperature of 10 K, as depicted in Fig. 1(b). The measurements were carried out within a section of the high-symmetry $\Gamma GZPX$ plane of the 3D Brillouin zone (BZ), illustrated in light blue in Fig. 1(a). By varying photon energies between 50 and 126 eV, different $k_z$ values were accessed, encompassing a substantial portion of the complete BZ. The presence of intense dispersionless Yb $4f$ states in close proximity to the Fermi energy ($E_F$) presented challenges in obtaining a clear FS. In Fig. 1(b), the photoemission intensity map at an energy of -0.3 eV reveals that the density of states is predominantly contributed to by conduction band electrons at this binding energy, with minimal input from the $4f$ electrons. Nonetheless, the FS sheets exhibited prominent 3D characteristics, with their shapes and intensities exhibiting distinct variations as a function of photon energy. A discernible FS sheet with clear periodicity was observed, facilitating the estimation of an inner potential ($V_0 \approx 13$ eV) and the determination of $k_z$ values for each photon energy.

In addition to the photon energy-dependent approach, constant photon energy measurements in the $k_x$-$k_y$ plane were employed to map the FS topology. Figures 1(c) and 1(d) illustrate the constant-energy contours of the band structure at a binding energy of -0.2 eV, recorded at 10 K with photon energies of 101 and 120 eV, respectively. The large FS sheet encompassing the BZ corners $\overline{X}$ displayed similarities at both photon energies, as shown in Fig. 2(a), indicative of an electron-like Fermi pocket. In contrast, the shape of the FS sheet centered at $\overline{\Gamma}$ exhibited significant variations with changing photon energies.

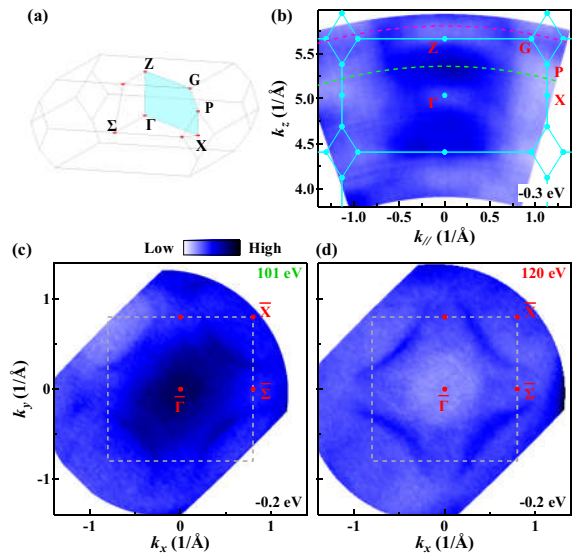

FIG. 1. (color online) YbCu$_2$Si$_2$ FS at low temperature. (**a**) A 3D BZ of YbCu$_2$Si$_2$ with high-symmetry momentum points marked by red dots. The $\Gamma GZPX$ plane in the bulk BZ is highlighted in cyan. (**b**) Experimental 3D FS maps measured using $h\nu$ = 50-126 eV photons in 1 eV steps, in the $\Gamma GZPX$ plane, integrated over an [-5 meV, 5 meV] energy windows with respect to the $E_F$. (**c**) and (**d**) Photoemission intensity maps integrated over an [-5 meV, 5 meV] energy windows with respect to the $E_F$ as a function of 2D momentum ($k_x$, $k_y$) taken with 101 and 120 eV phonons, respectively. The dashed gray lines represent the surface BZ. Momentum cuts with 101 and 120 eV photons are marked with bashed green and red lines, respectively, in (b).

The investigation of the valence band structure of Yb $4f$ electrons was carried out using a photon energy of 220 eV. At this energy, the photoionization cross-section of the Yb $4f$ state notably surpasses that of the Cu $3d$ state [31]. Figure 2(a) illustrates dispersion maps along the $\overline{X}$-$\overline{\Gamma}$-$\overline{X}$ direction over a temperature range spanning 10 to 280 K. In line with the complex nature of YbCu$_2$Si$_2$'s non-integer valence, both Yb$^{3+}$ ($4f^{13} \rightarrow 4f^{12}$) and Yb$^{2+}$ ($4f^{14} \rightarrow 4f^{13}$) contributions are evident in the photoemission spectrum [19, 32]. Throughout the measured temperature range, multiple flat bands and distinct valence bands are consistently observed.

As indicated by the red bars in Fig. 2(b), the $4f^{13} \rightarrow 4f^{12}$ multiplet, spanning from -12 to -5.5 eV, comprises at least nine levels originating from spin-orbit splitting. The multiplet near the $E_F$ results from the $4f^{14} \rightarrow 4f^{13}$ transition. Due to spin-orbit coupling, the final state is a doublet, denoted by the green bar, a feature commonly observed in other Yb-based HF systems [19, 33, 34]. The state approximately 1.3 eV below $E_F$ is attributed to the $4f_{5/2}$ state, while the state close to $E_F$ is ascribed to the $4f_{7/2}$ [33].

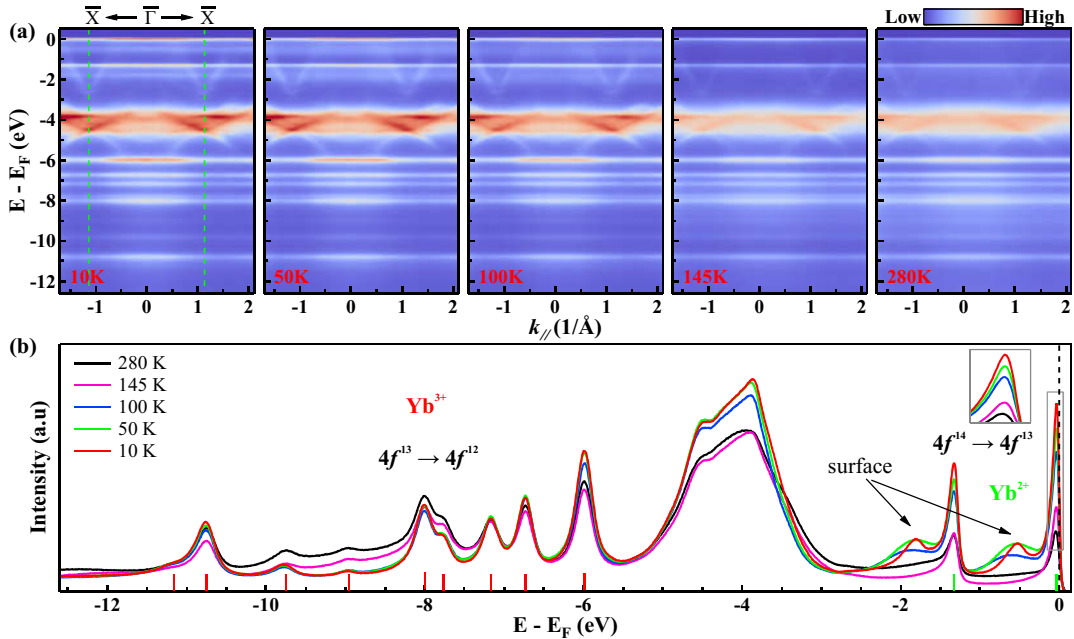

FIG. 2. (color online) Temperature dependence of the broad range electronic structure of $YbCu_2Si_2$. (a) ARPES data along $\overline{X}$-$\overline{\Gamma}$-$\overline{X}$ direction at the labeled temperatures. The momentum cut taken with 220 eV photons crosses $k_z \sim 0.3 \times 2\pi/c$. (b) Angle-integrated photoemission spectroscopy of the intensity plot in (a). The $Yb^{3+}$ ($4f^{13} \to 4f^{12}$) multiplet indicated by the red bars. And the bulk $Yb^{2+}$ ($4f^{14} \to 4f^{13}$) doublet indicated by the green bars. Two additional peaks were originated from the surface layer.

Remarkably, even at 280 K, well above the $T_K$, a distinct flat band remains prominently visible near $E_F$. This observation aligns with similar findings in other HF systems like Ce-based [5–10], U-based [11], and Yb-based [34] HFs, suggesting that Kondo screening processes are active even at elevated temperatures. Previous studies on $YbCu_2Si_2$ have demonstrated a gradual increase and decrease in the intensity of the $4f^{12}$ and $4f^{13}$ states, respectively, as temperature decreases [19]. The inset of Fig. 2(b) reveals that the intensity of the $4f_{7/2}$ state increases as temperature decreases, although the intensity of other $4f$ states do not vary monotonically. This observation may be attributed to the notable contribution of other orbitals, such as Cu $3d$ states, at the photon energy of 220 eV, which impedes the acquisition of a pure Yb $4f$ electron state. As shown in Figs. 2(a) and 2(b), the high-intensity dispersive conduction bands between -5 and -3 eV are primarily attributed to the Cu $3d$ states.

In addition to the prominent features, two broad, weaker peaks are observed at lower temperatures, emanating from the surface Yb layers [17, 19, 32]. At 10 K, these peaks are situated approximately 0.53 and 1.81 eV below $E_F$. However, their limited intensity implies a dominance of Si surface termination within the sample [17]. Furthermore, it is notable that the surface states exhibit a tendency to shift towards higher binding energies and broaden as temperature rises, ultimately vanishing at a higher temperature of $T \geq 145$ K.

Figures 3(a) and 3(b) display the temperature-dependent ARPES measurements along the $\overline{X}$-$\overline{\Gamma}$-$\overline{X}$ direction with 120 and 101 eV photons, respectively. The selection of these photon energies was strategic, aiming to enhance the photoconduction matrix element for Yb $4f$ and Cu $3d$ electrons, while effectively suppressing other electronic states [31]. Consequently, the spectra at these photon energies are predominantly characterized by dispersionless bands originating from Yb $4f$ states and dispersive conduction bands arising from Cu $3d$ states due to the influence of the photoconduction matrix element. This setup allows for the observation of $c$-$f$ hybridization processes.

Notable features in Figs. 3(a) and 3(b) are the significant flat bands and strongly dispersive conduction bands. The two prominent, almost flat bands stem from the spin-orbit splitting of the bulk $Yb^{2+}$ $4f$ final states. It is evident that these light conduction bands extensively hybridize with the flat bands, leading to strong momentum-dependent spectral intensities across all $f$-states. This observation strongly implies that Yb $4f$ electrons play a substantial role in forming the FSs [35]. The distribution of spectral weight differs with photon energy; for 120 eV, the spectral weight is most pronounced near the $\overline{X}$ point [Fig. 3(a)], while for 101 eV, it is strongest around the $\overline{\Gamma}$ point [Fig. 3(b)].

As the temperature decreases, a noticeable intensifica-tion of the $f$-bands and conduction bands is observed, highlighting the growing significance of $c$-$f$ hybridization. Intriguingly, even at 280 K, a temperature well above the $T_K$, the spectral weights of the $f$-bands exhibit discernible momentum dependence, signifying the commencement of $c$-$f$ hybridization at this elevated temperature.

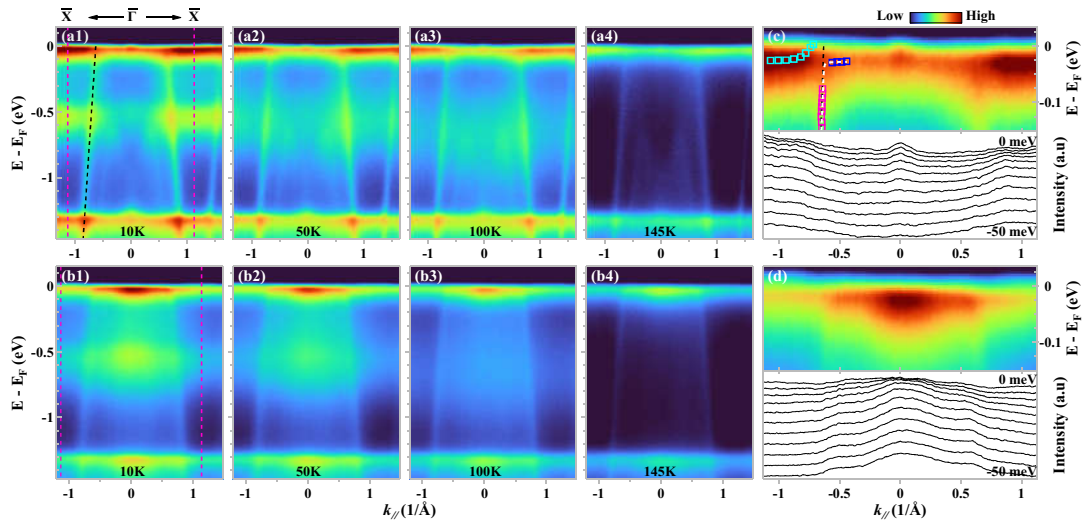

FIG. 3. (color online) Temperature evolution of the electronic structure of YbCu$_2$Si$_2$. (a) and (b) ARPES spectra along the $\overline{X}$-$\overline{\Gamma}$-$\overline{X}$ direction at the labeling temperature with 120 and 101 eV photons, respectively. (c) Top: zoom-in view of the quasiparticle dispersion at 10 K near $E_F$ of (a1). Bottom: momentum distribution curves as a function of binding energy. The empty symbols represent quasiparticle dispersions. The hybridized dispersion was fitted by the equation based on the PAM, as indicated by the white-dashed line. The black-dashed line represents the bare conduction band. (d) Top: zoom-in view of the quasiparticle dispersion at 10 K near $E_F$ of (b1). Bottom: momentum distribution curves as a function of binding energy.

Figures 3(c) and 3(d) zoom into the vicinity of the electronic structure near the $E_F$ at 10 K for 120 and 101 eV, respectively. Clear $c$-$f$ hybridization can be observed with a small indirect hybridization energy gap, which can be described by a periodic Anderson model (PAM) [36, 37]. As shown by the cyan squares, the hybridization leads to a very shallow dispersion, forming an electron-like pocket around the $\overline{X}$-point. It is consistent with theoretical calculations based on $4f$-itinerant LDA+$U$ method [21]. Consistent with theoretical predictions [21], substantial structural features were detected at the $\overline{\Gamma}$ point. To gain a comprehensive understanding of these features, higher energy resolution measurements are imperative.

As shown in Figs. 2 and 3, the surface-derived low energy flat (SLF) band of YbCu$_2$Si$_2$ located approximately -0.53 eV exhibits strong signals at 10 K but gradually weakens with increasing temperature, ultimately vanishing at 145 K. An additional observation is the shift of the SLF band towards higher binding energies as temperature rises. In contrast, the positions of the bulk-contributed $4f$ and conduction bands remain relatively stable.

During ARPES measurements, surfaces are often subject to aging due to the adsorption of residual molecules. This aging can result in the degradation or even disappearance of surface state signals due to the reduction in photoelectrons' mean free path. In light of this, we conducted systematic investigations to trace the evolution of the SLF band with respect to both time and temperature.

Figures 4(a) and 4(b) present a systematic investigation into the temporal and thermal evolution of the peak position of the SLF band, respectively. Sample A, subjected to cleavage and measurement at a low temper-ature of 10 K, exhibited minimal variation in the SLF band's peak position across different photon energies, aligning harmoniously with the anticipated behavior of a surface state. Significantly, the SLF band in sample A remained discernible for an extended period exceeding 15 hours post-cleavage, accompanied by only subtle positional shifts. These nuanced shifts hint at the delicacy of any energy perturbation affecting the SLF band, possibly stemming from adsorption of residual molecules.

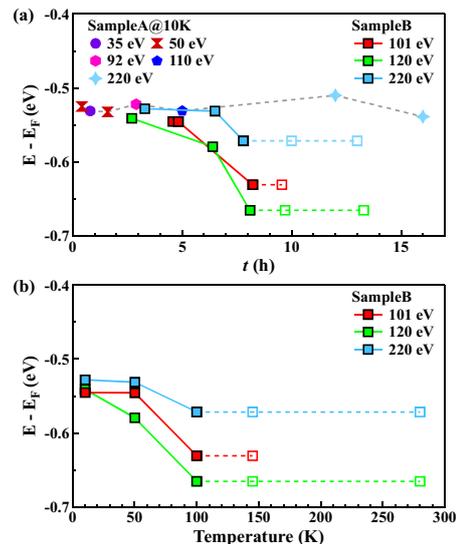

FIG. 4. (color online) Time- and temperature-dependent of the surface Yb layer derived LS band. (**a**) The binding energy of the LS peak is plotted against time from cleaving. (**b**) The binding energy of the LS peak is plotted as a function of temperature. The unfilled markers indicate that the LS band is not observable.

In marked contrast, sample B, cleaved at 10 K and subsequently subjected to heating for measurements, unveiled pronounced alterations. As the temperature ascended to 145 K, the SLF band became imperceptible for the three employed photon energies, as denoted by the unmarked regions in Fig. 4(b). The question arises: What underlies the vanishing of the SLF band at elevated temperatures? It is manifest that surface aging resulting from gas adsorption alone cannot satisfactorily account for the disappearance of the SLF band, as it persisted in sample A for a considerable period surpassing 15 hours post-cleavage. A distinct hypothesis suggests the absence of Yb atoms in the topmost surface layer as a plausible explanation. However, preceding investigations on YbB$_{12}$, as conducted by Hagiwara et al., have convincingly demonstrated that Yb atoms endure on the surface even after enduring annealing temperatures as high as 1400 K [20]. Consequently, the hypothesis of surface Yb atom absence appears incongruous with the measured temperatures.

In light of these findings, we contemplate that the temperature may have induced modifications at the surface of YbCu$_2$Si$_2$. To account for this unexpected behavior, we consider that the effective Kondo temperature at the surface differs significantly from that in the bulk. Two plausible mechanisms may contribute to this scenario:(i) Surface Crystal Structure Relaxation: Reduced atomic coordination at the surface leads to weakened $c$-$f$ hybridization [38, 39]. A distinctive variance between the characteristic temperatures at the surface and in the bulk has been observed in Ce-based HF systems [16, 40]. However, in contrast to our observations, the characteristic temperature in CeRh$_2$Si$_2$ is considerably higher at the surface than in the bulk, contrary to expectations arising from reduced coordination and diminished hybridization at the surface [16]. (ii) Valence State Transition: With the elevation of temperature, a transition in the valence state of surface Yb atoms, specifically from Yb$^{2+}$ to Yb$^{3+}$ valence states, assumes plausibility as an explanatory mechanism. This proposition aligns with the observations in Fig. 2(b). Owing to the strong 4$f$ hybridization, Yb-based HF systems often exhibit mixed valency at the surface [18].

Another noteworthy change is observed as the temperature increases: the SLF band shifts towards higher binding energies while the bulk bands remain relatively stable, as evident in Figs. 2 and 3. Notably, for 120 eV photons, the SLF peak exhibits a shift of over 100 meV as the temperature rises from 10 K to 100 K. In general, surface states can undergo shifts in response to the adsorption of residual gases. To address this, a comparison of the results for samples A and B was conducted, excluding the possibility that the shift was a consequence of an extended adsorption period, resembling an overlayer-like behavior. While Pauli repulsion or changes in the surface work function may offer explanations for the shift of the SLF band, they do not account for its complete disappearance at higher temperatures.

Collectively, the observations lead us to consider that the behavior of the SLF band is likely a result of multiple factors. These factors may encompass electron transfer between the surface state and bulk atoms/adsorbates, coupled with the influence of Pauli repulsion and potential adjustments in the surface work function.

## V. CONCLUSIONS

In summary, our study involved a comprehensive exploration of the temperature-dependent surface and bulk Yb 4$f$ spectra within the Kondo lattice YbCu$_2$Si$_2$ using ARPES over a wide temperature range. The key findings of our investigation offer compelling evidence for two main conclusions: (i) Yb 4$f$ electrons initiate their transition into the formation of Heavy Fermion states at temperatures considerably higher than the Kondo Temperature $T_K$; (ii) Our research reveals a notable disparity in the effective Kondo temperatures between the surface and bulk regions, with the effective Kondo temperature of the surface being markedly lower than that of the bulk. This outcome aligns with the theoretical anticipation of reduced coordination and, consequently, weaker hybridization at the surface. Collectively, our achievements in this study establish the feasibility of quantitatively distinguishing the distinct contributions of surface and bulk in the Kondo processes. This capacity opens avenues for testing the theory of many-body interactions, thereby advancing our understanding of complex electronic behaviors in heavy fermion systems and related systems.

## ACKNOWLEDGMENTS

This work was supported by the National Natural Science Foundation of China (Grant No. 12074436), the National Key Research and Development Program of China (Grant No. 2022YFA1604204), and the Science and Technology Innovation Program of Hunan Province (2022RC3068). We are grateful for resources from the High Performance Computing Center of Central South University.

* Corresponding author: jqmeng@csu.edu.cn